\documentclass[conference]{IEEEtran}
\IEEEoverridecommandlockouts
\usepackage{cite}
\usepackage{amsmath,amssymb,amsfonts}
\usepackage{algorithmic}
\usepackage{graphicx}
\usepackage{textcomp}
\usepackage{xcolor}
\usepackage{array}
\def\BibTeX{{\rm B\kern-.05em{\sc i\kern-.025em b}\kern-.08em
    T\kern-.1667em\lower.7ex\hbox{E}\kern-.125emX}}
\begin{document}



\begin{titlepage}


       \vspace*{1cm}

       \copyright2024 IEEE. Personal use of this material is permitted.  Permission from IEEE must be obtained for all other uses, in any current or future media, including reprinting/republishing this material for advertising or promotional purposes, creating new collective works, for resale or redistribution to servers or lists, or reuse of any copyrighted component of this work in other works.

       \vspace{1.5cm}
  
\end{titlepage}



\title{Design Frameworks for Spatial Zone Agents \\ in XRI Metaverse Smart Environments

\thanks{Tri-council of Canada, Canada Research Chairs program.}
}

\author{\IEEEauthorblockN{Jie Guan}
\IEEEauthorblockA{\textit{Adaptive Context Environments Lab} \\
\textit{OCAD University}\\
Toronto, Canada \\
jie.guan@ocadu.ca}
\and
\IEEEauthorblockN{Jiamin Liu}
\IEEEauthorblockA{\textit{Adaptive Context Environments Lab} \\
\textit{OCAD University}\\
Toronto, Canada \\
jiaminliu@ocadu.ca}
\and
\IEEEauthorblockN{Alexis Morris}
\IEEEauthorblockA{\textit{Adaptive Context Environments Lab} \\
\textit{OCAD University}\\
Toronto, Canada \\
amorris@ocadu.ca}

}


\maketitle

\begin{abstract}
The spatial XR-IoT (XRI) Zone Agents concept combines Extended Reality (XR), the Internet of Things (IoT), and spatial computing concepts to create hyper-connected spaces for metaverse applications; envisioning space as zones that are social, smart, scalable, expressive, and agent-based. These zone agents serve as applications and agents (partners, assistants, or guides) for users co-living and co-operating together in a shared spatial context. The zone agent concept is toward reducing the gap between the physical environment (space) and the classical two-dimensional user interface, through space-based interactions for future metaverse applications. This integration aims to enrich user engagement with their environments through intuitive and immersive experiences and pave the way for innovative human-machine interaction in smart spaces.
Contributions include: i) a theoretical framework for creating XRI zone/space-agents using Mixed-Reality Agents (MiRAs) and XRI theory, ii) agent and scene design for spatial zone agents, and iii) prototype and user interaction design scenario concepts for human-to-space agent relationships in an early immersive smart-space application.

\end{abstract}

\begin{IEEEkeywords}
Virtual Reality, Mixed Reality, Augmented Reality, Internet of Things, Human-Computer Interaction, Metaverse
\end{IEEEkeywords}

\section{Introduction}



The Metaverse refers to the merger of computing technologies that blend virtual and physical spaces; applying technologies such as Artificial intelligence and Extended Reality and the internet-of-things \cite{lee2021all}. These technologies aim to provide digital twins of the real world as well as completely virtual shared environments; however, in both cases there is a clear gap between the user's physical environment and their virtual environment. This is referred to as the metaverse disconnect problem and researchers have explored methods to increase the connection between users and their relationships to smart spaces \cite{Guan2022ExtendedMetaverseIEEE,Guan2023ThesisCross-Reality,Guan2023ThesisPoster,Guan2022ExtendingThesis}. 
To enhance the communication and connection between the metaverse and physical environments and objects, the method of XRI (XR-IoT)\cite{Morris2021XRIWorkstation,Tsang2021XRITaxonomy} is introduced for supporting the hyper-connected metaverse environment. These XRI components allow for the creation of hybrid physical-virtual objects as well as the hybrid physical-virtual interactions that can take place. As more ``hybrid objects'' are combined into a user's space, this results in a ``hyper-connected'' environment that can be social, smart, engaging, and immersive, as in \cite{guan2023socialxri}. Such environments are poised to enrich the user's information architecture, as in \cite{resmini2023being} in terms of placemaking (presence), consistency(mental models), resilience (adaptation), reduction (context adaptation), and correlation (exploration), across applications and smart space configurations (small home spaces, work spaces, larger scale city environments, etc).

\begin{figure}[ht]
 \centering 
 \includegraphics[width=1\linewidth]{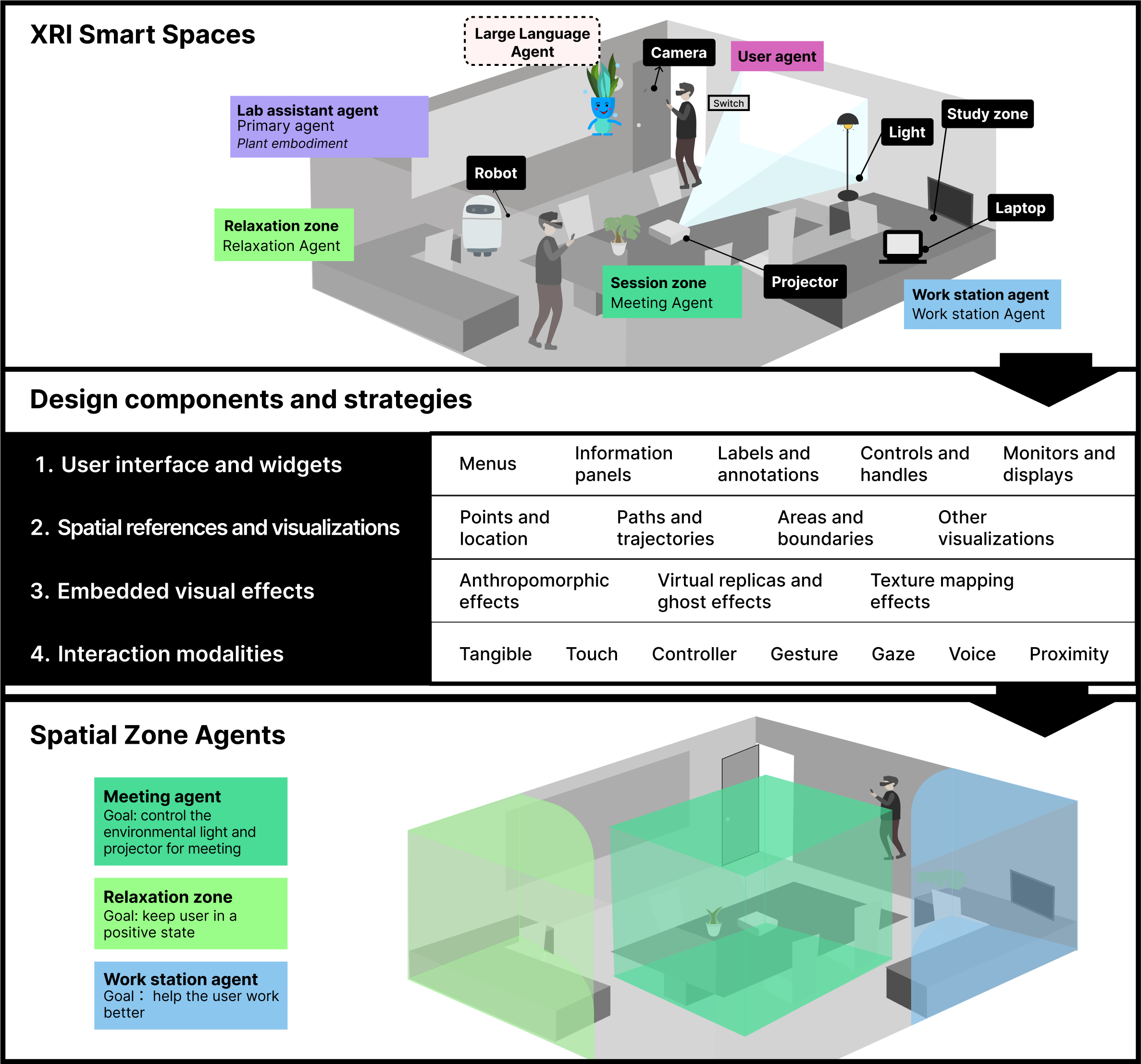}
 \caption{XRI smart-spaces \cite{Morris2021XRIWorkstation} can be transformed into zone-aware spatial agents that have potential to interact with users based on their spatial contexts. This work focuses on design components and strategies for these agents \cite{suzuki2022augmented-robot}.}
 \label{Introduction}
\end{figure}


However, these designs still have interaction challenges between humans, agents, and spaces of the metaverse context. This vision of the metaverse requires user, multi-agent, XRI, and Spatial Computing Interfaces in order to make the hybrid environment fit the user application context. Figure \ref{Introduction} presents a high-level XRI smart space (objects, IoT components, cameras, as well as virtual agents, etc), and highlights that multiple design components and strategies for mixed reality interactions, from \cite{suzuki2022augmented-robot}, can be considered toward not just mixed reality or IoT objects, but also spatial zone agents (or zone as agent designs). This work presents the design of example zone-based agents and agent interactions using an early framework. It addresses the question of how to design a spatial zone agent for metaverse interactions, and an approach for creating zone agent prototypes. Contributions include: i) a theoretical framework for creating XRI zone/space-agents using Mixed Reality Agents (MiRAs) theory \cite{holz2011mira} and XRI, ii) agent and scene design for spatial zone agents, and iii) prototype and user interaction design scenario and concepts for human-to-space agent relationships in an early immersive smart-space application. 

\begin{figure}[ht]
 \centering 
 \includegraphics[width=\linewidth]{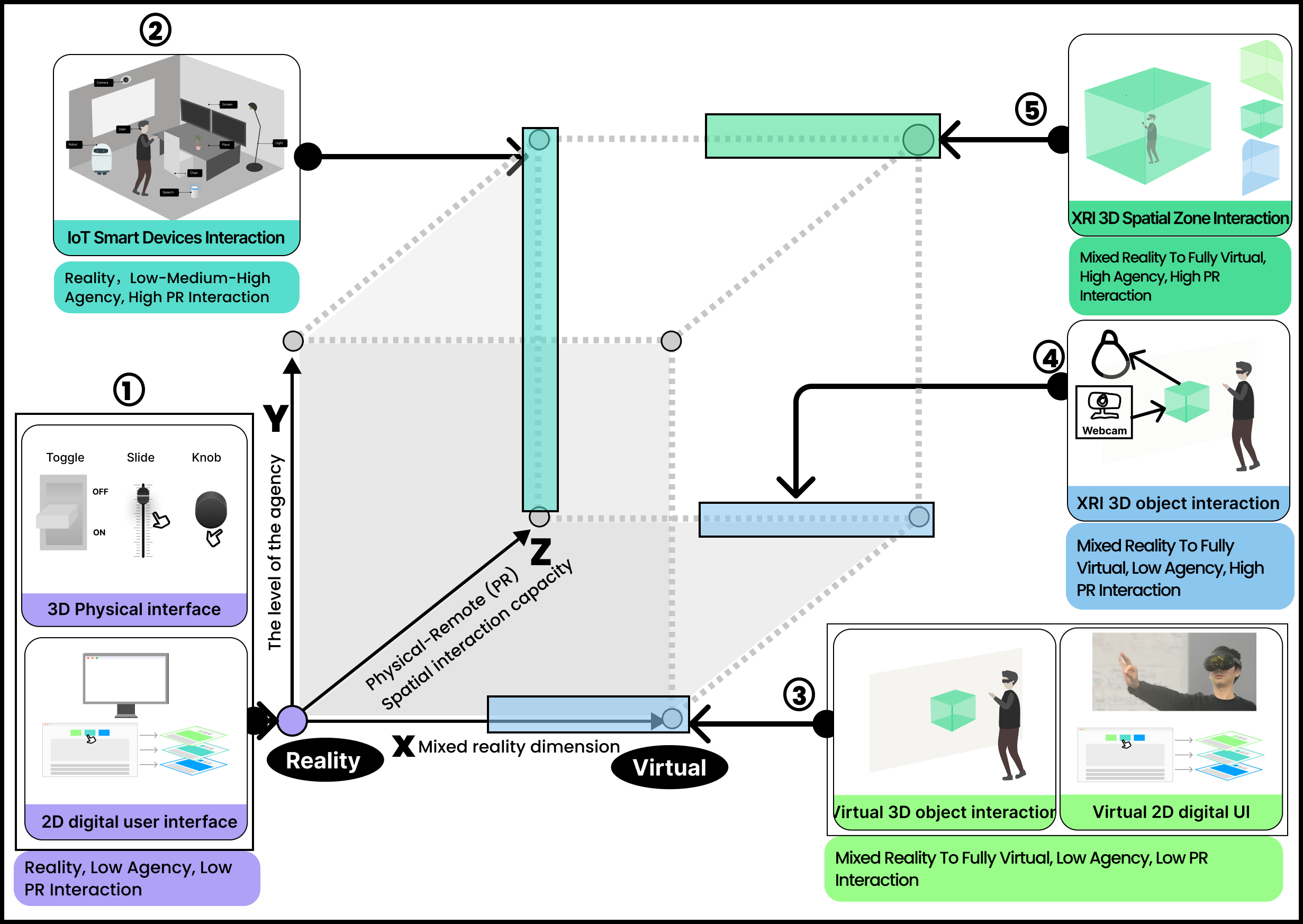}
 \caption{Zone-aware spatial agents can have a range of Physical and Remote interaction abilities and properties, as well as Mixed Reality Agency (MiRAs), as seen in examples 1-5, based on \cite{holz2011mira}\cite{suzuki2022augmented-robot}.}
 \label{Theory}
\end{figure}
\section{Background}

\textbf{XRI and Metaverse:} 
The early frameworks and proof-of-concept prototypes of XR and IoT for specific scenarios had been presented in \cite{jo2016ariot}\cite{jo2019ar}  to connect the physical information and the virtual GUI and objects. Based on this foundation, previous works also explored a virtual plant embodied in Mixed Reality and connected to the physical environment via IoT in mobile device\cite{Shao2019IoTAvatar}, and an enhanced head-mounted display prototype, in \cite{Guan2020IoTAvatarVideo}\cite{Morris2020IoTAvatar2}. The XRI concept also appears in \cite{Tsang2021XRITaxonomy} with a taxonomy and workstation prototype in \cite{Morris2021XRIWorkstation}.   

Metaverse has diverse definitions as it is a developing term in the state of the art. As in \cite{dionisio20133d}, the metaverse is considered to be constructed by the immersive and three-dimensional digital online environments, while in\cite{lee2021all}\cite{ning2021survey}, it is considered metaverse is constructed by internet, web technologies and extended reality with hybridization of physical and virtual space. Metaverse has been considered as the next generation of the internet in both industry and academia \cite {cheng2022will}.
Together, the metaverse as a virtual online environment is envisioned to increase the connectedness with the physical space. Hence, the concept of extending the metaverse with XRI has been presented in previous works such as in \cite{Guan2022ExtendingThesis,Guan2022ExtendedMetaverseIEEE,Guan2023ThesisPoster,Guan2023ThesisCross-Reality,guan2023socialxri}.

\textbf{Spatial Computing and interface:} Spatial computing incorporates the concept and technologies to create a new understanding of locations, improving the relationship between humans and space \cite{Shekhar_SpatialComputing}. Spatial Computing is the human-in-the-loop interaction that engages with real objects and space; together with the recent trend of mixed reality that merges real and virtual worlds, the need for mapping virtual environments in the real-world is becoming more important\cite{greenwold2003spatial}.
Spatial computing allows users to interact freely with the 2D interface and three-dimensional information that spatial augmented reality provides \cite{marner2014spatial}.

\textbf{Agent System Design (Agency):} 
Agent System Design is an interdisciplinary field, within Artificial Intelligence, as in\cite{wooldridge1995intelligent}, which has been instrumental in laying the groundwork for creating intelligent and scalable agent-based systems. This foundation has opened avenues for innovation in various realms, including Mixed Reality. The Mixed Reality Agents (MiRAs)\cite{holz2011mira} framework considers agents that are designed to operate within mixed reality environments, with dimensions in embodiment, interaction and agency level. Moreover, the robot as an agent has been explored within the domain of Human-Robot Interaction (HRI) and Extended Reality (XR)\cite{suzuki2022augmented-robot}, where they can act to bridge the chasm between the human environment and the digital world \cite{szafir2019mediating}.

Agent systems integrate AI, software design, and HCI \cite{jennings1998roadmap}, cognitive architectures, like Prometheus \cite{padgham2002prometheus}, and other beliefs-desires-intentions (BDI) frameworks, and machine learning models, like those used in robotics and games \cite{caillou2017simple,bordini2020agent}, and deploy approaches like Behavior Trees for decision-making \cite{kyaw2013unity,iovino2022survey}. Similarly, in entertainment, Emotional AR Agents can enhance immersion and interactivity \cite{ushida1998emotion, gushima2017ambient, bylieva2019virtual}, and in HRI, agent embodiment can influence user perception, particularly with recent advancements in generative AI (like DALL-E and VQGAN) revolutionizing agent visual representation and understanding capabilities \cite{wang2019influence, groom2009evaluating, ramesh2021zero, crowson2022vqgan}.

Together, these indicate new directions for future metaverse spaces that are agent-based, zone-oriented, and immersive.

\begin{figure}[ht]
 \centering 
 \includegraphics[width=.9\linewidth]{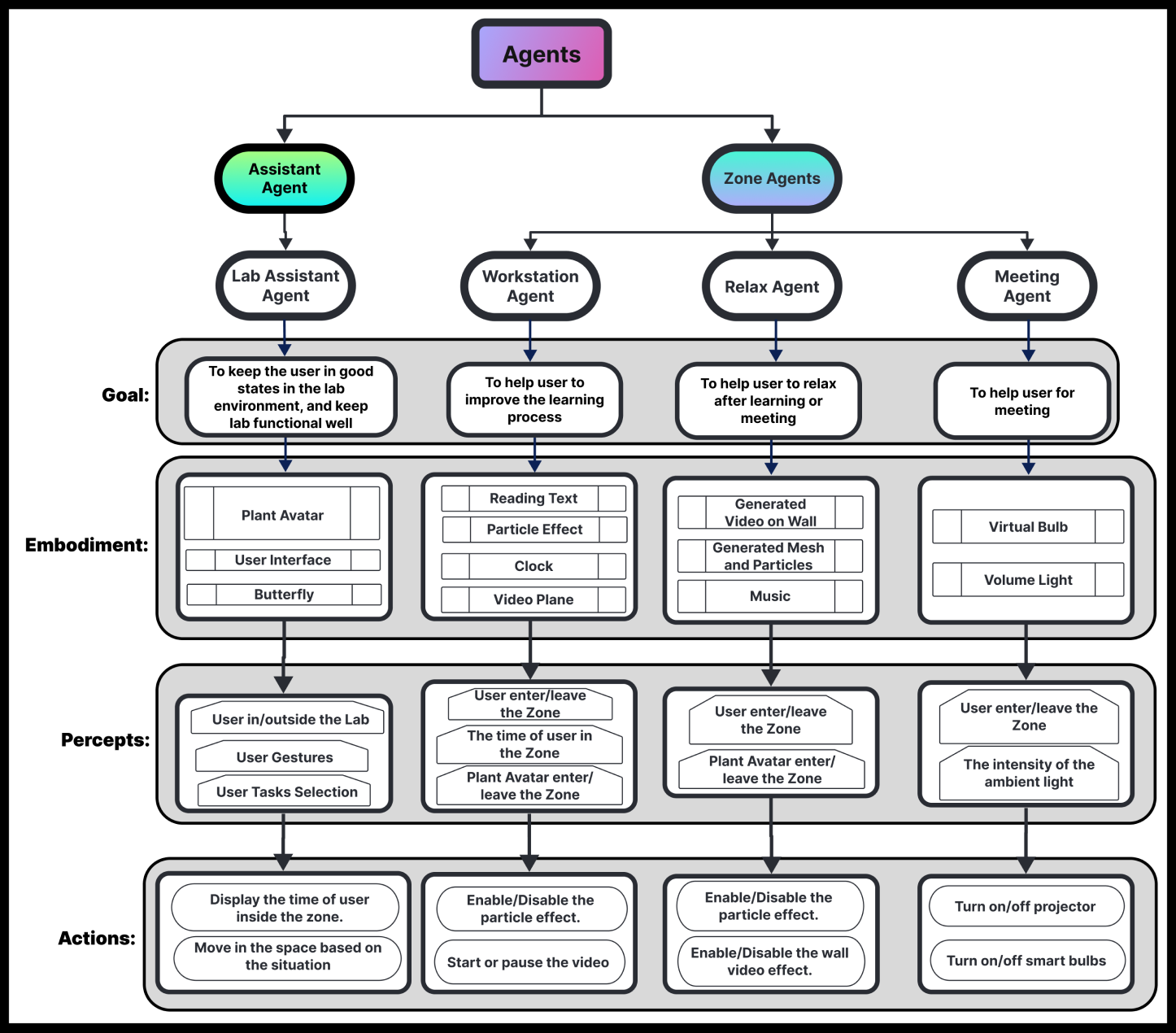}
 \caption{Zone agents are here designed for spatial context, based on agent design methodology like \cite{padgham2002prometheus}\cite{wooldridge1995intelligent}, (including monitoring and responding to user-zone events such as time context, virtual context, IoT context, zone object context, etc.). These agents express embodiment through zone-driven actions that may be virtual or physical. The level of agency can vary according to agent implementation (from simple reflex agents to high-functioning agents). }
 \label{AgentAndScene}
\end{figure}

\begin{figure*}[ht]
 \centering 
 \includegraphics[width=0.8\linewidth]{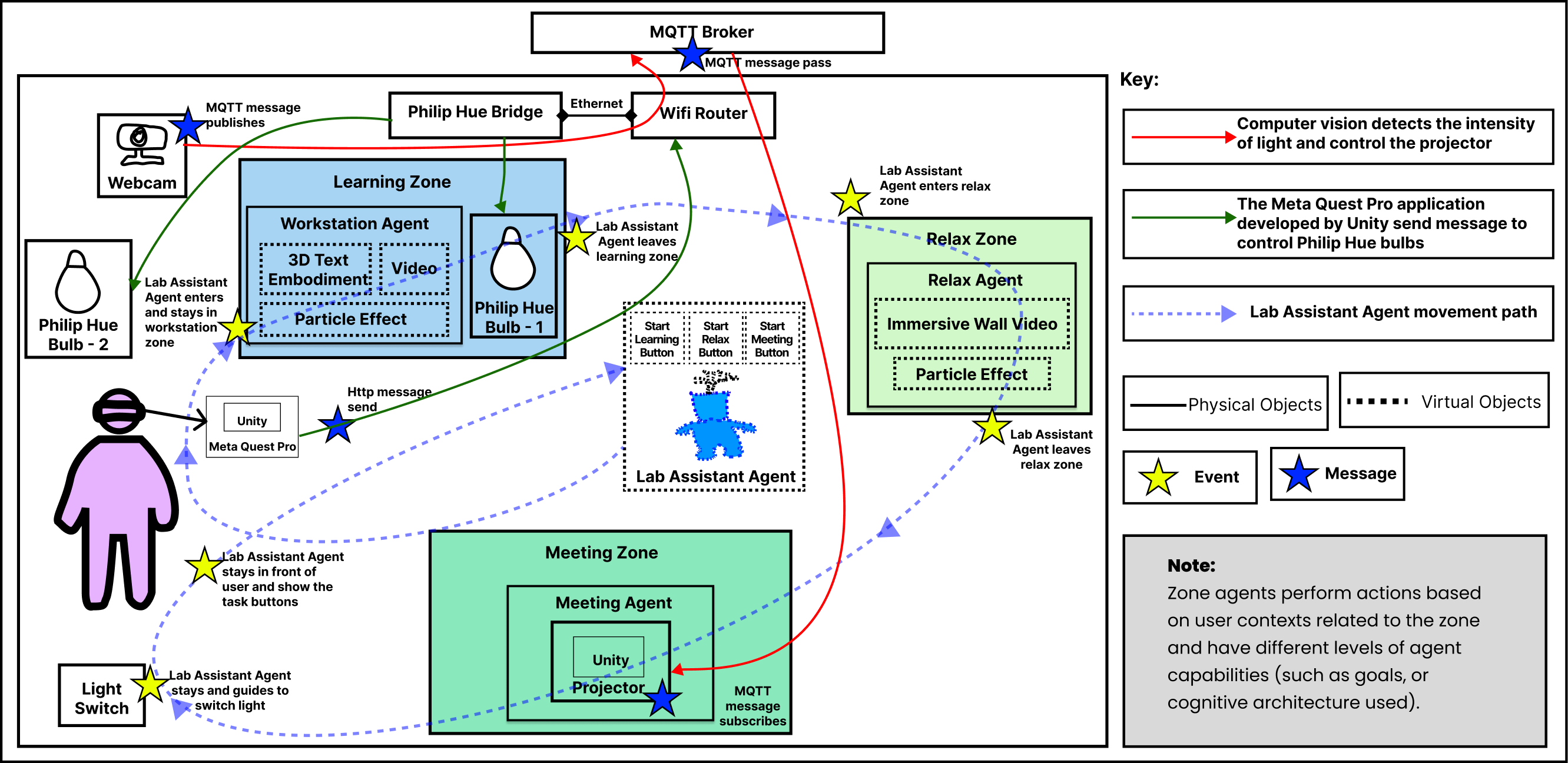}
 \caption{The prototype framework for the multi-agent Zone scenario includes the highlighted actors (user, learning agent, relax agent, meeting agent); these agents interact with the user; which may be both physical, IoT, or virtual interactions, across a communication channel (e.g., MQTT).} 
 \label{Framework}
\end{figure*}

\section{Theory - Designing a Spatial Zone Agent for XRI Smart Spaces}

\begin{figure*}[ht]
 \centering 
 \includegraphics[width=.8\linewidth]{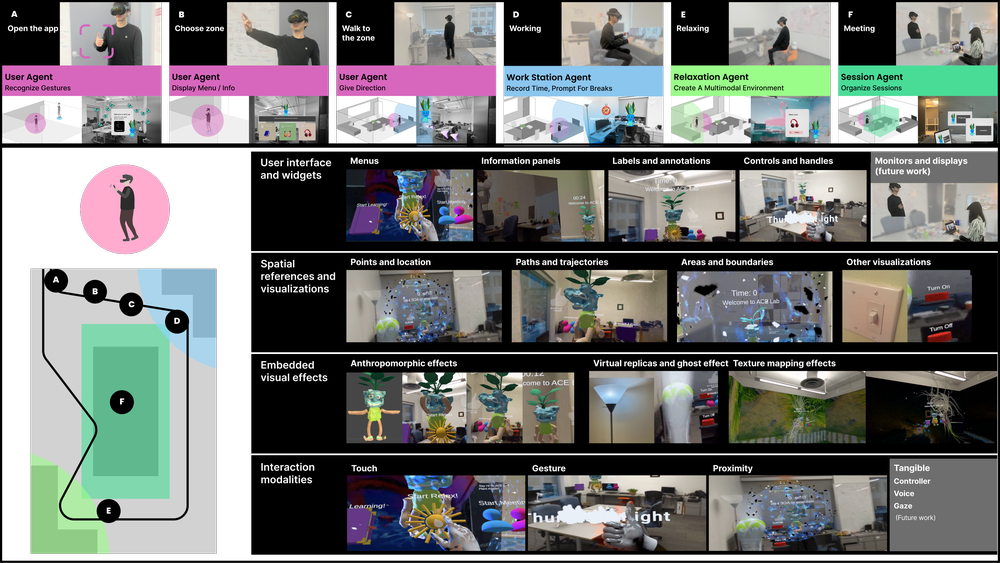}
 \caption{Using the HRI Framework \cite{suzuki2022augmented-robot}, the design of Zone agents can be considered; showing user interface and widgets, spatial references and visualizations, embedded visual effects, and interaction modalities for an early prototype, as shown. A deeper evaluation of zone agents is left for future work.} 
 \label{Prototype}
\end{figure*}

\textbf{Design Dimensions for Spatial Zone Agents: MiRAs and XRI and HRI Taxonomy\cite{suzuki2022augmented-robot}}
Figure \ref{Theory} presents the XRI Zone Agent design theory, which has been merged and adapted from XRI theory, virtuality continuum \cite{milgram1994taxonomy}, MiRAs\cite{holz2011mira} and human-robot interaction \cite{suzuki2022augmented-robot} into three dimensions, including the mixed reality dimension, the level of agency, and physical-remote (PR) spatial interaction capacity. The X axis is the mixed reality dimension adapted from the virtuality continuum \cite{milgram1994taxonomy} that represents the real-environment on the left and fully virtual reality environment on the right. The Y axis is the level of agency, which means how capable the agent is in terms of its autonomous, reactive, proactiveness, and social abilities \cite{wooldridge1995intelligent}. The Z axis of physical-to-remote (PR) spatial interaction capacity represents the level of physical or remote control users have when interacting with objects, and ranges from physical interaction to non-physical interaction, such as through IoT communication.

On the left side, the \textit{number 1} represents physical interaction and interface design, with the 3D physical interfaces (the physical toggle, slider, etc.), empowering users to interact with and dynamically adjust elements within the physical environment, and a 2D digital user interface (screen icons are widely used for human-machine interaction). The \textit{number 2} indicates the IoT-enabled physical objects and integrated controllers, and this setup facilitates the communication of information across a network of devices. In addition, autonomous agents, such as robots, have the capacity to function within and adapt to this environment while establishing interaction with users. 

On the right side, the \textit{number 3} is from mixed reality to fully virtual environments, where users manipulate 2D digital UI and 3D objects. Current mixed-reality interactions commonly employ 3D virtual objects with hand-tracking capabilities to simulate interaction with real-world objects. The \textit{number 4} presents that the virtual objects could be used to control or be manipulated by physical objects through IoT communication, as in \cite{Guan2022ExtendedMetaverseIEEE}\cite{Guan2023ThesisCross-Reality}\cite{guan2023socialxri}. Finally, the \textit{number 5} shows XRI 3D Spatial Zone Interaction. These agents and their design capabilities and objectives are shown in Figure \ref{AgentAndScene}.


The scenario considers the designed spatial position of different agents (with virtual embodiment) and the potential IoT-Enabled devices that could be controlled (such as a light, projector, and robot). In terms of spatial zone agents these can include a relaxation agent, meeting agent, and workstation agent, each operating as applications within the space, driven by specific objectives and responding to user contexts.

\section{Design Framework and Prototype of Spatial Zone Agents}


\textbf{XRI Zone Agent Framework:} Figure \ref{Framework} shows the framework of the prototype, indicating how the user navigates the use case environment (lab environment) between zones and the interaction with both virtual and physical objects through IoT communication. The virtual agents such as lab assistant agent are rendering in XR headset, powered by Unity\footnote{https://unity.com/}, and the physical objects, such as Phillips Hue smart lights\footnote{https://www.philips-hue.com/en-ca/p/hue-white-and-color-ambiance-a19---e26-smart-bulb---75-w--2-pack-/046677563370} and Webcam are communicating to the virtual agents through an MQTT Broker\footnote{https://mqtt.org/}.

\textbf{Prototype of XRI Zone Agent Interactions:}
Figure \ref{Prototype} presents the prototype of the XRI Zone Agent, for the scenario and user behaviour interaction within the lab environment with the XRI zone agents. When users wear the XR HMD device and enter the lab environment, they can see the virtual lab assistant agent represented as a plant avatar, with butterflies moving in the environment. The avatar is a guide that shows welcome messages to the user and displays an initial introduction to the user. There is also an indicator to tell the user how to interact and enable the next state of the plant avatar with a thumbs-up gesture (hand tracking). When the user shows the thumbs-up gesture (see Figure \ref{Prototype}), the plant avatar will wave and move toward the user and show three menu icons of tasks (start learning, start relaxing, and start meeting) for selection that could be performed in the lab. When the user presses the start learning button, for example, the plant avatar will move toward the learning zone with the workstation agent and show an indicator on that zone to guide the user to the location (this is an example of a spatial agent that can move between zones). The user will enter the ``learning zone'', and the workstation agent for that zone will show the time of how long the user has been studying (this is an example of tracking user context in the zone). Once the time reaches the desired length, the plant avatar will stand in front of the user to ask if the user would like to start relaxing. If the user presses the start relaxing button, the plant avatar will guide the user toward the relax zone, and the wall around the user, driven by the relax agent, will transform into a relaxed scene with particle effects (an example of a simple-reflex zone agent based on user entry). Also, when the user wants to start a meeting, they can perform the thumbs-up gesture, to summon the plant avatar and select the meeting state. The plant avatar will move to the physical light switch to ask/guide the user to turn it off. Once the meeting zone agent detects the environment light has turned off (through computer vision light detection), the meeting projector will turn on and enter the meeting mode (another example of zone agent responsiveness). While these agent interactions are simple-reflex agents, they can be enhanced by integration of stronger agent paradigms, such as the use of ML model-driven controllers, large language models, or more capable cognitive architectures.

\section{Summary}
 
        



This work has provided a theoretical framework for considering smart-space zones as agents, and has proposed an early design-theoretic exploration of mixed reality and IoT (XRI) zone agents as well as an architecture and initial prototype implementation. This has considered that user interactions with zone/space agents will involve sensors, actuators, embodiment, decision-making, and a mixture of communication paradigms. The work considers exploring space-to-human and human-to-space interactions driven-by mixed-reality agents and user and environmental context within zones. It envisions that zone agents would be designed to adapt to user spatial contexts and needs through changes in either physical or virtual reality. A high-level user interaction design has been presented, based on three zone agents for work, leisure, and meetings. The hope is that such framing of space-as-agent will help further new forms of metaverse exploration where digital twins of everyday environments work together with humans-in-the-loop, streamlining their shared contexts. This early exploration sets the stage for further research in this direction, and future research will refine further in terms of agency level, physical-virtual responsiveness, and new forms of spatial interaction.

\section*{Acknowledgment}
 This work gratefully acknowledges funding from the Tri-council of Canada under the Canada Research Chairs program.

\bibliographystyle{IEEEtran}
\bibliography{references}

\end{document}